\input{aipcheck}

\documentclass[
    ,final            
  ]
  {aipproc}

\layoutstyle{8x11single}
\usepackage{amsmath,amssymb}      
\usepackage{aas_macros}

\begin{document}

\title{Search for neutrinos from Gamma-Ray Bursts with ANTARES}


\author{Eleonora Presani}{
  address={Nikhef,  Science Park 105,  1098 XG Amsterdam}
}

\begin{abstract}
A method to search for neutrino induced showers from gamma-ray bursts in the ANTARES detector is presented.  ANTARES consists of a three-dimensional array of photosensitive devices that measure Cherenkov light induced by charged particles produced by high energy neutrinos interacting in the detector vicinity. The shower channel is complementary to the more commonly used upgoing muon channel. The corresponding detection volume is smaller, but has the advantage of being sensitive to neutrinos of any flavour.
\end{abstract}

\maketitle
\newcommand{\x}{\times}
\newcommand{\X}{$\times$}
\newcommand{\ra}{\rightarrow}
\newcommand{\Ra}{$\rightarrow$}
\newcommand{\circa}{$\sim$}
\newcommand{\prop}{\simeq}
\newcommand{\sun}{\odot}
\newcommand{\E}[1]{\x 10^{#1}}
\newcommand{\tx}[1]{\textrm{#1}}
\newcommand{\de}{\partial}
\newcommand{\e}{\textrm{e}}
\newcommand{\pa}{_\parallel}
\newcommand{\pe}{_\perp}
\newcommand{\susy}[1]{\tilde{#1}}
\newcommand{\gradi}{^{\circ}}

\newcommand{\fr}{\vec}            
\newcommand{\h}{\hat}            
\newcommand{\bma}[1]{\hbox{$\,{\bf{#1}}$}}   
\newcommand{\p}{\cdot}           
\newcommand{\vp}{\times}          
\newcommand{\sI}{\hspace{0.3cm}}
\newcommand{\sII}{\hspace{0.6cm}}
\newcommand{\sIII}{\hspace{0.9cm}}
\newcommand{\sd}{\hspace{1.2cm}}
\newcommand{\sva}{\vspace{0.5cm}}
\newcommand{\svb}{\vspace{1.0cm}}
\newcommand{\svc}{\vspace{1.5cm}}
\newcommand{\svd}{\vspace{2.0cm}}
\newcommand{\pv}{\sa;\sa}

\newcommand{\units}[1]{\hbox{$\,{\rm{#1}}$}}            
\renewcommand{\u}[1]{\hbox{${\rm #1}$}}                
\newcommand{\s}{\,\u{s}}
\newcommand{\ms}{\,\u{ms}}
\newcommand{\mus}{\,\mu\u{s}}
\newcommand{\ns}{\,\u{ns}}
\renewcommand{\d}{\,\u{d}}
\newcommand{\y}{\,\u{y}}
\newcommand{\sr}{\,\u{sr}}
\newcommand{\mum}{\,\mu\u{m}}
\newcommand{\mm}{\,\u{mm}}
\newcommand{\cm}{\,\u{cm}}
\newcommand{\m}{\,\u{m}}
\newcommand{\km}{\,\u{km}}
\newcommand{\pc}{\,\u{pc}}
\newcommand{\kpc}{\,\u{kpc}}
\newcommand{\Mpc}{\,\u{Mpc}}
\newcommand{\ly}{\,\u{ly}}
\newcommand{\Mly}{\,\u{Mly}}
\newcommand{\UA}{\,\u{U\, A}}
\newcommand{\kg}{\,\u{kg}}
\newcommand{\gr}{\,\u{g}}
\newcommand{\eV}{\,\u{eV}}
\newcommand{\keV}{\,\u{keV}}
\newcommand{\MeV}{\,\u{MeV}}
\newcommand{\GeV}{\,\u{GeV}}
\newcommand{\TeV}{\,\u{TeV}}
\newcommand{\PeV}{\,\u{PeV}}
\newcommand{\EeV}{\,\u{EeV}}
\newcommand{\ZeV}{\,\u{ZeV}}
\newcommand{\eVsq}{\,\u{eV^2}}
\newcommand{\pbinv}{\,\u{pb^{-1}}}
\newcommand{\G}{\,\u{G}}                                   
\newcommand{\muG}{\,\mu\u{G}}                              
\newcommand{\T}{\,\u{T}}                                   
\newcommand{\N}{\,\u{N}}                                   
\newcommand{\W}{\,\u{W}}                                   
\newcommand{\K}{\,\u{K}}                                   
\newcommand{\GHz}{\,\u{GHz}}
\newcommand{\MHz}{\,\u{MHz}}
\newcommand{\kHz}{\,\u{kHz}}
\newcommand{\srad}{\,\u{srad}}
\newcommand{\erg}{\,\u{erg}}

\newcommand{\nue}{$\nu_{e}$}
\newcommand{\num}{$\nu_{\mu}$}
\newcommand{\nut}{$\nu_{\tau}$} 
\newcommand{\pota}{$^{40}\tx{K }$}
\newcommand{\anti}[1]{\overline{#1}}


\newcommand{\riga}{\vspace{0.5cm}}


\section{Introduction}

Through this analysis, the Antares experiment is sensitive to the electromagnetic and hadronic showers induced in the detector by the interaction of neutrinos originating from Gamma-Ray Bursts. 

In the Fireball model, neutrinos are produced during the GRB explosion via photo-pion production in collision of ultrarelativistic protons with  photons in the jet of the GRB. Although the production mechanism favours the muon neutrino it can safely be assumed, due to neutrino oscillation,  that the flavour ratio at Earth is $\nu_{e}:\nu_{\mu}:\nu_{\tau} = 1:1:1$. 

While the standard track search is sensitive only to muon neutrinos, interacting via charged current interaction with nuclei, a shower analysis is sensitive to all flavours. In order to deal with the high background introduced by atmospheric muons, a track search is generally limited to upgoing neutrinos, i.e. neutrinos that traversed the Earth and interacted near the detector. The current shower analysis takes advantage of the time coincidence with a triggered GRB to reduce the background and can therefore extend the field of view to the entire solid angle. 
To estimate the discovery potential and the sensitivity of the detector, a Waxman-Bahcall flux, averaged for a single GRB, was assumed. The analysis consists of the search for coincidences between well reconstructed showers in Antares and photons triggered in one of the satellites which subscribed to the IPN (Interplanetary Gamma-Ray Burst Timing Network).

\section{Neutrino Production in GRBs}

In the  Fireball model \cite{Meszaros2006} charged particles can be accelerated to high energies via the Fermi acceleration mechanism.
Neutrinos are primarily produced via the  $p+\gamma$ (photo-hadronic) process. The Waxman-Bahcall flux for a single average Gamma-Ray Burst \cite{waxman1997} was used for the shape and normalization of the flux distribution. It is a diffuse flux, and it is therefore necessary to derive the flux of a single average GRB. This procedure relies on a number of assumptions: the average duration for each GRB is given by T90=50~s and 667 GRBs are, on average, detected per year by satellites. This number is based on the observations of the CGRO mission. The resulting flux can be written as:
\begin{equation}
J_{WB}^{single} = J_{WB}^{diff} \p 4 \pi \frac{N_{sec}^{year}}{T_{90}^{avg}667}
\end{equation}
where the $4\pi$ takes into account the solid angle in which the flux is generated, $N_{sec}^{year}$ is the number of seconds in a year, and the denominator represents the average number of GRB multiplied by their average duration (chosen to be 50~s).
\section{The Antares Detector and  Event Signatures}

The Antares (Astronomy with a Neutrino Telescope and Abyss environmental RESearch) neutrino telescope \cite{ANTDet}  is located at around 40 km off Toulon, at a depth of 2475 m.   The photosensors are arranged along a length of $\sim 350$m of vertical cable. A total of 885 optical sensors are carried by 12 flexible lines. Each line comprises up to 25 detection storeys, each equipped with three downward-looking 10-inch photomultipliers (PMTs), oriented at $45^{\circ}$ relative to the vertical. Each PMT is housed in an Optical Module (OM) that consists of a 17-inch glass sphere in which the optical connection between the PMT and the glass is ensured by an optical gel.

The majority of neutrinos that reach the Earth will just pass through it. However, it is possible for them to undergo a weak interaction with a nucleon. Due to the extremely small cross section, a very large target mass is necessary to attempt the detection. The reaction that is used in high energy neutrino detectors to reveal this particle is the neutrino deep inelastic scattering (DIS)  with a matter nucleon.  If a  $W^{\pm}$ boson is exchanged (charged current interaction), a charged lepton and an hadronic shower are generated, while if a $Z^0$ boson is exchanged (neutral current) an hadronic shower is followed by another neutrino.
\begin{figure}[h]
\includegraphics[width=0.3\linewidth]{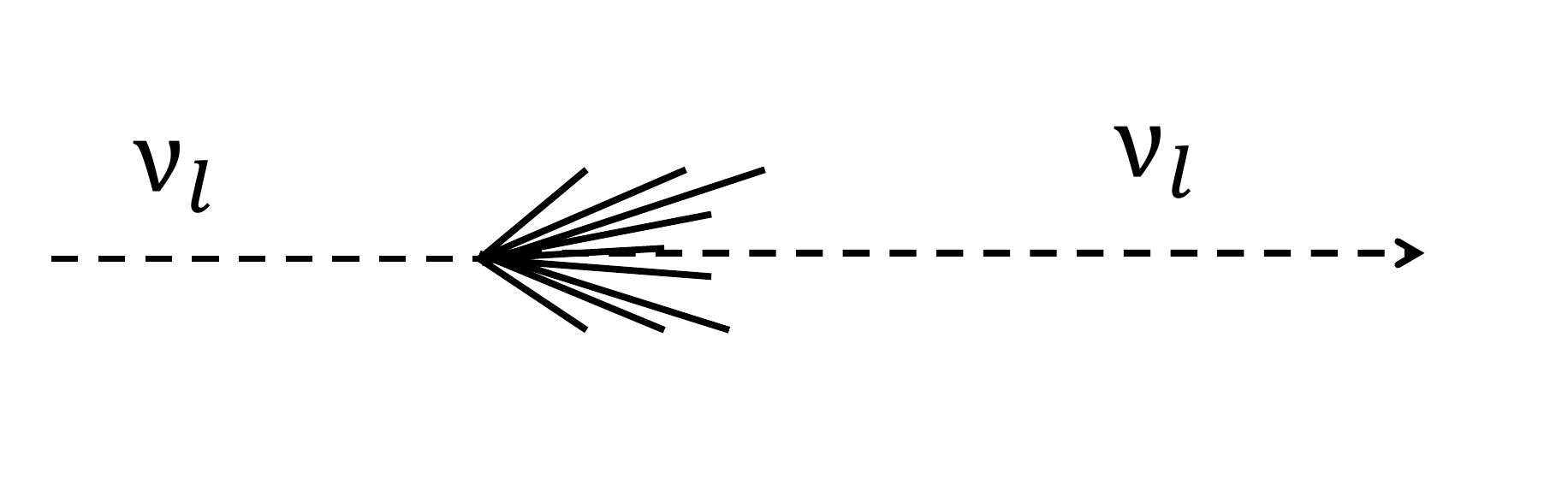}
\includegraphics[width=0.3\linewidth]{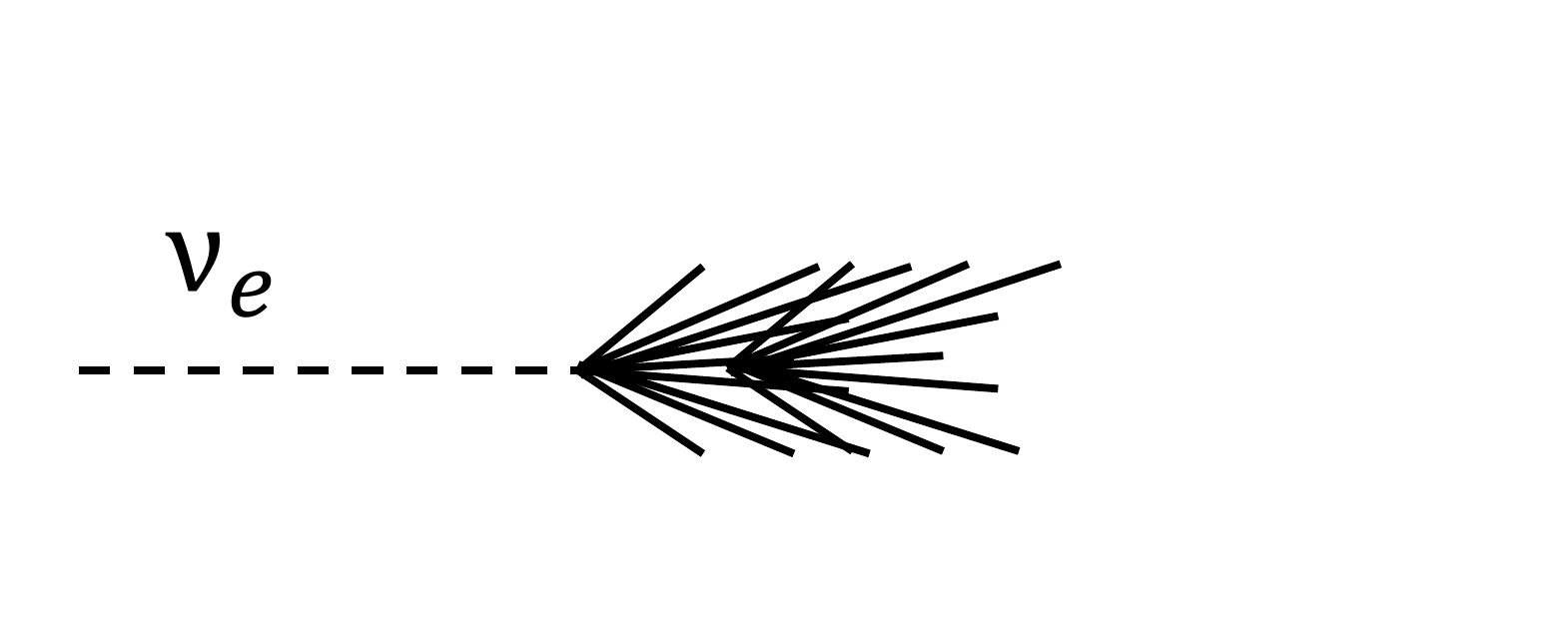}
\caption{Schematic view of a  ``shower'' like event in the Antares detector. Neutrinos of any flavour will produce an hadronic shower in the detector when interacting via NC (left). Electron neutrinos will produce a shower in the detector for both NC and CC interactions (right).}%
\label{fig:nushower_signatures}
\end{figure}
 When a neutrino interacts in the vicinity of the detector via a charged current (CC) interaction, it generates a relativistic charged particle that will travel through the sea water in the detector. When the speed of a charged particle exceeds the velocity of light in the medium through which it is travelling it will induce the emission of a radiation called Cherenkov radiation \cite{Leo1988}. 
The light emission creates a wavefront where the emitted light is coherent. The wavefront forms a cone with its apex at the travelling particle. The opening half-angle of the cone $\theta_c$ can be written as $ \cos \theta_c = 1/\beta n$.
The sea water at the location of the ANTARES neutrino telescope has a value of n of about 1.35, thus the value of the Cherenkov angle is about $42.2^{\circ}$.  

Neutrinos of any flavour  interacting via a NC will produce a neutrino, which remains undetectable, and an hadronic shower, consisting of many interaction products. Due to the small volume where the shower takes place the shower appears in the detector as a point-like light source developing isotropically in time. Tau neutrinos can also be detected with shower analysis, as their interaction and decay produce showers. A neutrino that undergoes a NC interaction will not produce a charged lepton capable of leaving a detectable track (Fig. \ref{fig:nushower_signatures}). If the main vertex of the interaction is in the vicinity of the detector, it is possible to see light as if being emitted from a point source.  For an hadronic (or electromagnetic) shower to be measured, it is necessary for the primary vertex to occur in or around the instrumented detector volume. This reduces the probability of production of a reconstructible shower. 

\section{Shower Event Reconstruction}

The majority of the photons measured in Antares (called ``hits'') are due to optical background and to Cherenkov light induced by down-going muons created  in interactions of high energy cosmic rays in the atmosphere. For this work a strict hit selection is used to distinguish between background and signal hits.  The hit selection starts by searching for clusters of hits in time, because these are more likely to be due to a signal than background. During the hit selection, the detailed geometry of a single storey is ignored and all hits on a floor are considered together to form clusters. 
For each event, all hits on one storey are time ordered, all hits that occur within 20~ns of each other are merged together. Using these as seeds, all the hits that are causally connected to the clusters are kept. To be sure that only signal hits are selected,  only hits compatible with being generated by the same particle are selected \cite{Brunner2011}.
\begin{figure}[h]
\includegraphics[width=0.25\linewidth]{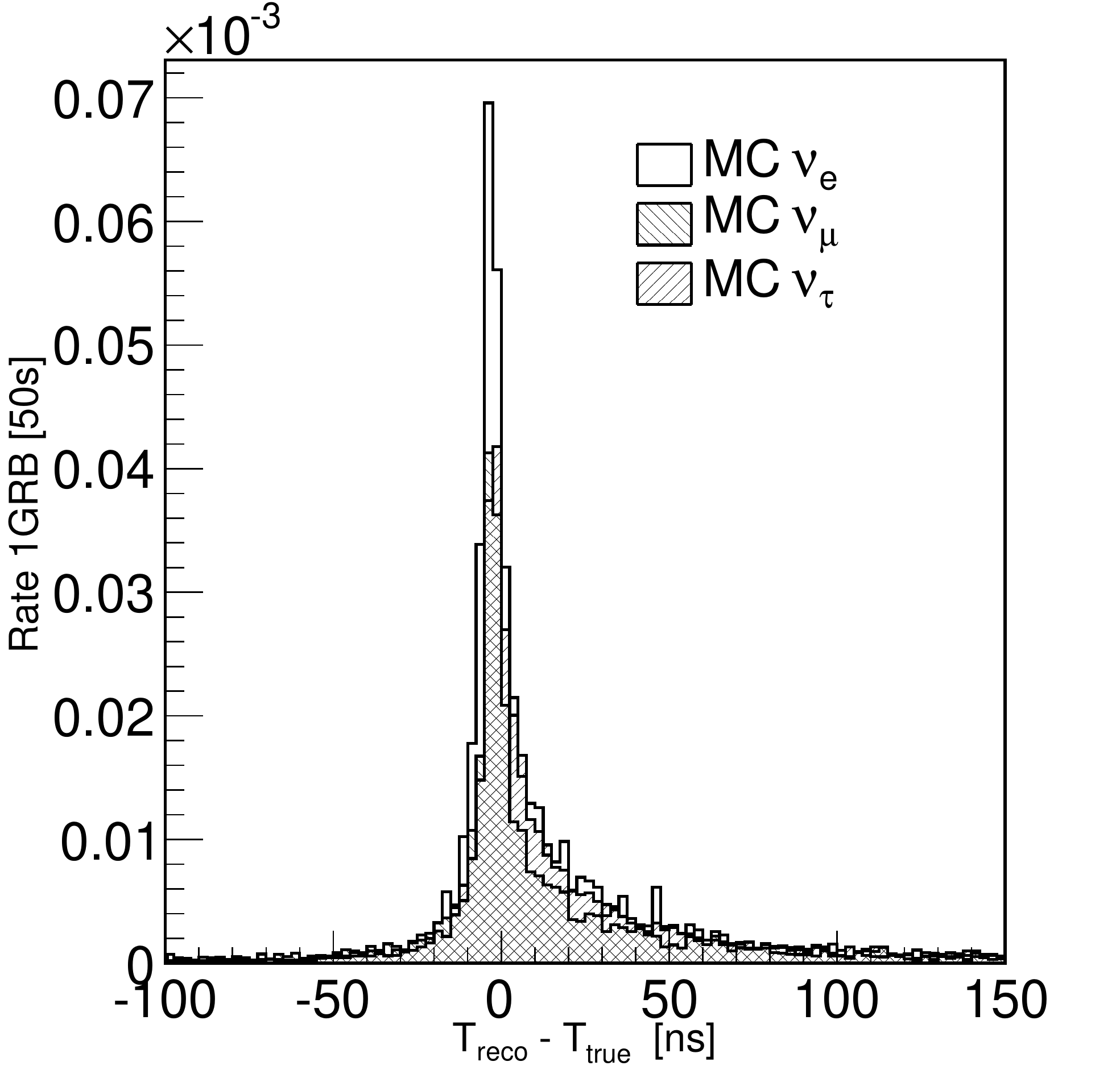}
\includegraphics[width=0.25\linewidth]{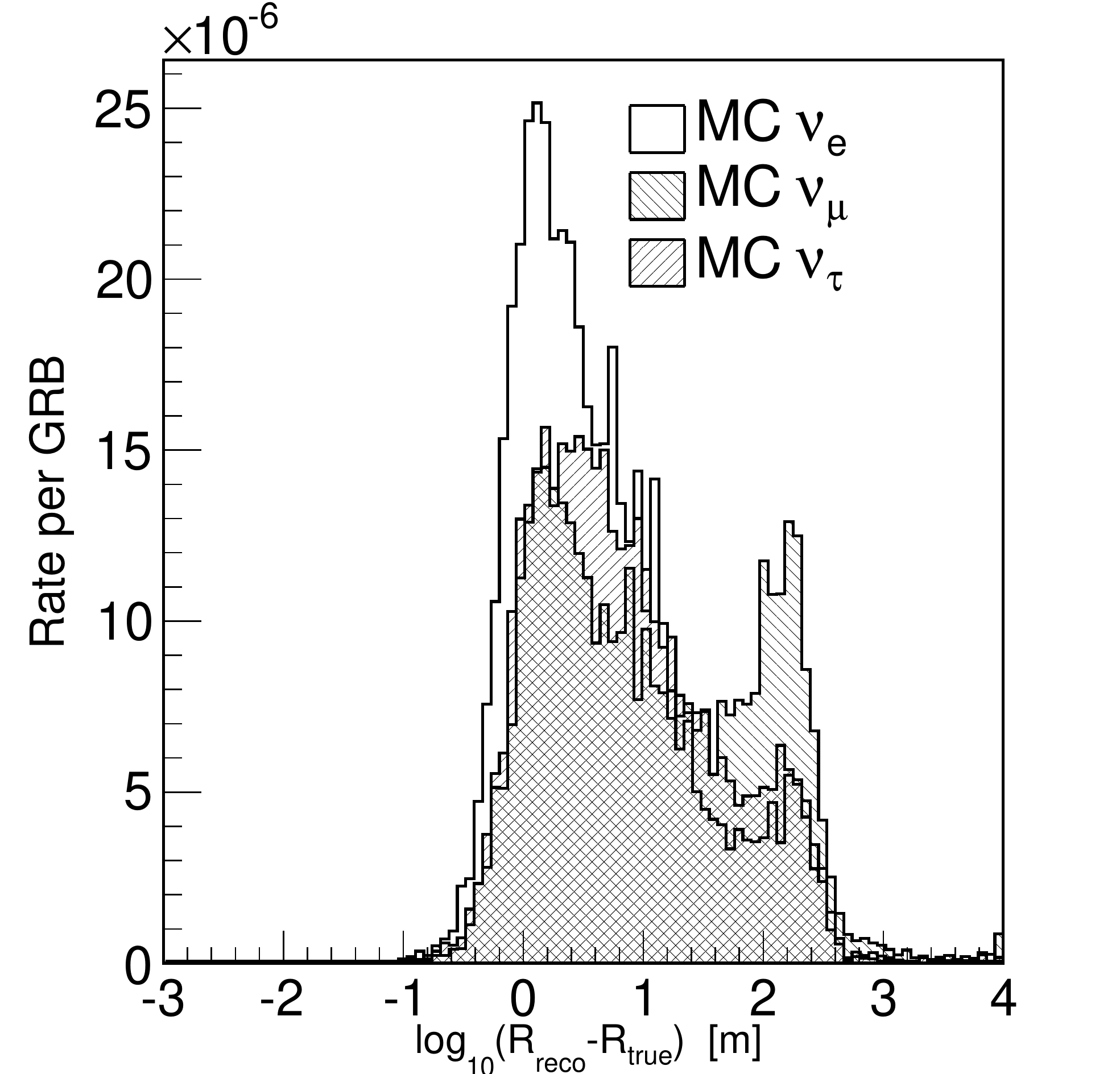}
\caption{Time and spatial resolution of the reconstruction. The first shows the difference between the reconstructed and the true time of the shower. The time resolution of the reconstruction taken to be the RMS of the distribution is 5.1 ns. The second shows the logarithm of the three dimensional distance between the true and reconstructed shower vertex. The spacial resolution is best for electron neutrinos, with a median value of 3.0 m. }\label{fig:reco_resolution}
\end{figure}
Given a shower occurring at a position $(X_{true},Y_{true},Z_{true})$ and time $t_{true}$ and emitting light, the expected arrival time, $t_{exp}^i$, of photons is given by Pythagoras' rule:
\begin{equation}
t_{exp}^i = t_{true} + \frac{n}{c}\sqrt{(X_i-X_{true})^2+(Y_i-Y_{true})^2+(Y_i-Y_{true})^2}\label{eq:reco_expectedtime}
\end{equation}
where $n$ is the group refractive index of sea water ($n = 1.3797$) and $c$ is the speed of light in vacuum. To reconstruct the time $t_{reco}$ and position $(X_{reco},Y_{reco},Z_{reco})$ of the shower, these unknown parameters are varied until the difference between the expected time and the measured time of each selected hit is minimised. The minimising function chosen for this analysis is the so called M-estimator function, defined as:
\begin{equation}
\rho(t_{meas}^i,t_{exp}^i) = M(t_{meas}^i,t_{exp}^i) = 2*\sqrt{1+\frac{(t_{meas}^i-t_{exp}^i)^2}{2\sigma^2_i}} - 2
\end{equation}\label{eq:reco_mest}
\noindent%
The errors $\sigma_i$ could in principle be different for each hit. In the Antares experiment, all optical modules are observed to have a similar resolution and, therefore, the same value of $\sigma_i = \sigma = 1~ns$ is used for each hit.

The resolution of the reconstruction is shown in Figure~\ref{fig:reco_resolution}, where spectrum is generated weighting the events with a Waxman-Bachall spectrum..  The time resolution is measured to be 5.1~ns. The right plot in  Figure~\ref{fig:reco_resolution} shows the three dimensional distance between the true and reconstructed shower. The spatial resolution of the reconstruction is given by the median of this distribution. It is better for electron neutrinos, with a value of 3.0 m, and worse for muon neutrinos, with a value of 9.3 m. The reason for this difference is that electron neutrinos always produce a shower-like signal. On the other hand, only muon neutrinos that undergo a NC interaction will produce a pure shower signature. The second peak, especially visible for muon neutrino induced events, is due to Bremsstrahlung and electromagnetic showers generated along the muon track that are reconstructed far away from the interaction vertex.
\begin{figure}[t]
\includegraphics[width=0.35\linewidth]{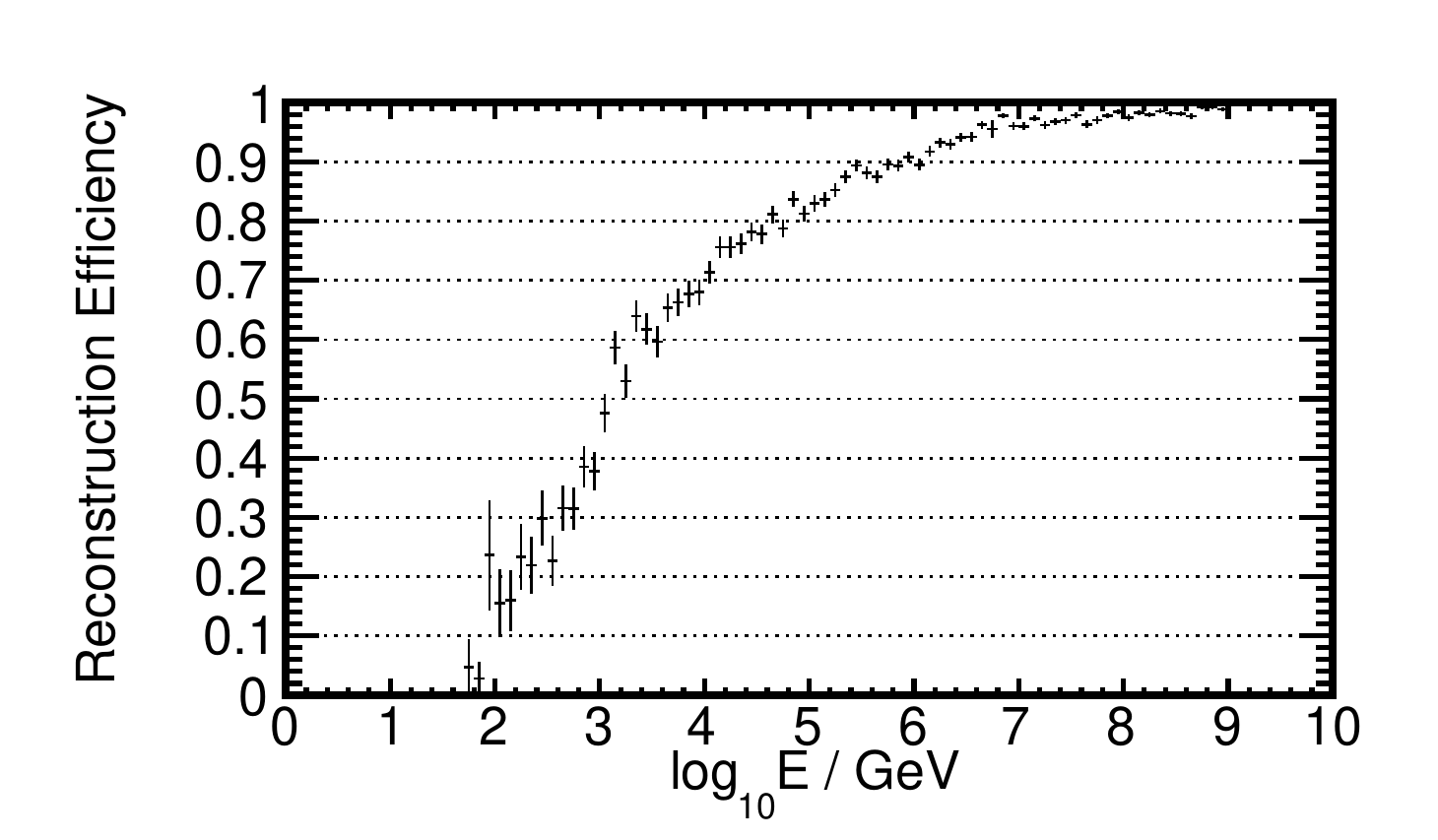}
\includegraphics[width=0.33\linewidth]{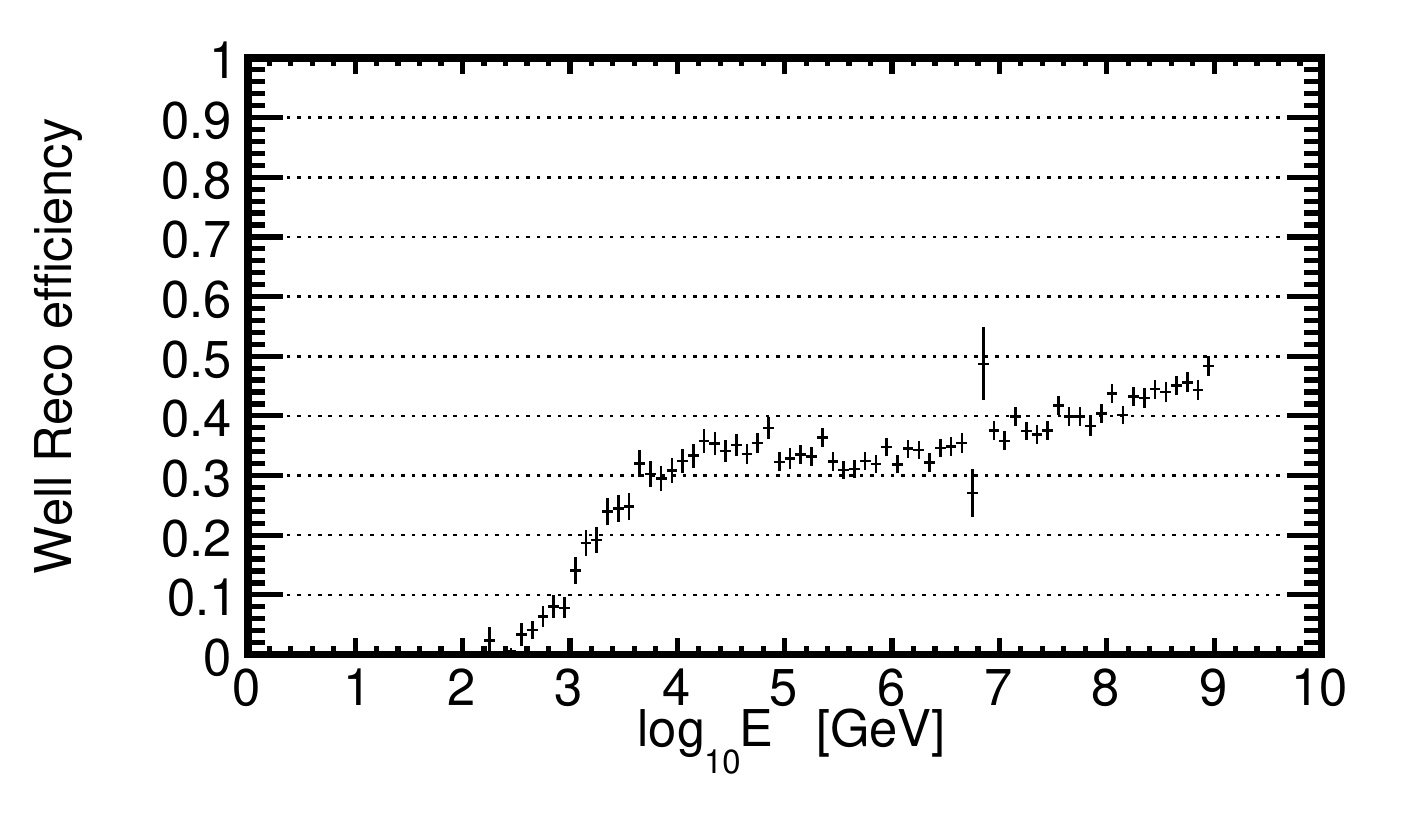}
\caption{Reconstruction efficiency as a function of neutrino energy for an average GRB flux. See text for details.}\label{fig:reco_efficiency}
\end{figure}
The efficiency of the reconstruction has also been studied. The efficiency of the shower reconstruction itself depends on the criteria used to select reconstructed showers. A minimum bias reconstruction efficiency can be determined by accepting all events in which the reconstruction successfully found a minimum. For this work, in addition, 5 selected hits on at least 2 lines were required for the minumum selection. The efficiency of the reconstruction as a function of the neutrino energy after this first selection is shown in the first plot of Figure~\ref{fig:reco_efficiency}. The reconstruction is considerably less efficient at low energies, where showers produce fainter light and less hits. At high energy practically all triggered events are reconstructed. The right plot in Figure~\ref{fig:reco_efficiency} shows the efficiency with which the shower fit procedure determines very accurately the position and time of the shower. Only events for which the reconstructed shower was within 3~m and 10~ns of the true shower vertex are used for calculating this efficiency. Even with these strict cuts on the quality of the reconstruction, the efficiency stays between 30\% and 50\%, especially in the energy range on which the analysis is focused.

\section{Neutrino Induced Shower Analysis}

The concept of time correlated analysis is very simple: given a certain search time window, one looks for a neutrino event in the same time window as the one of a gamma-ray burst event. Quality cuts on the reconstructed events are tuned in order to reduce the background and to optimize the discovery potential. Since this analysis is focused on shower events, no angular cut is applied, and the entire sky is considered. The main advantage of using a time correlated search is the efficient background rejection, due to the small time window considered.  The time window used for the quality cut optimization is 100 seconds. Each GRB will be analysed for the duration of its T90. A set of data corresponding to 295.8 days of data taking during the year 2008 has been used to extract the background level. During the same year 65 GRBs have been recorded and considered for this work.

The quality cut optimization was done on three parameters: the quality parameter (the M-estimator of the reconstruction),  the number of hits assumed to come from the shower and the number of lines used for the reconstruction of the shower.  It is important to have well-reconstructed shower, therefore the quality parameter is very significant. The hits coming directly form the shower are the set of all the selected hits that have a time residual smaller than 15~ns.  This parameter is important as it gives some information on how good the shower hypothesis is for that event. If the event does not behave as a shower, the hits will not be distributed as expected. The number of detector lines used in the reconstruction is correlated  with the energy of the shower.
\begin{figure}[h]
\includegraphics[width=0.45\linewidth]{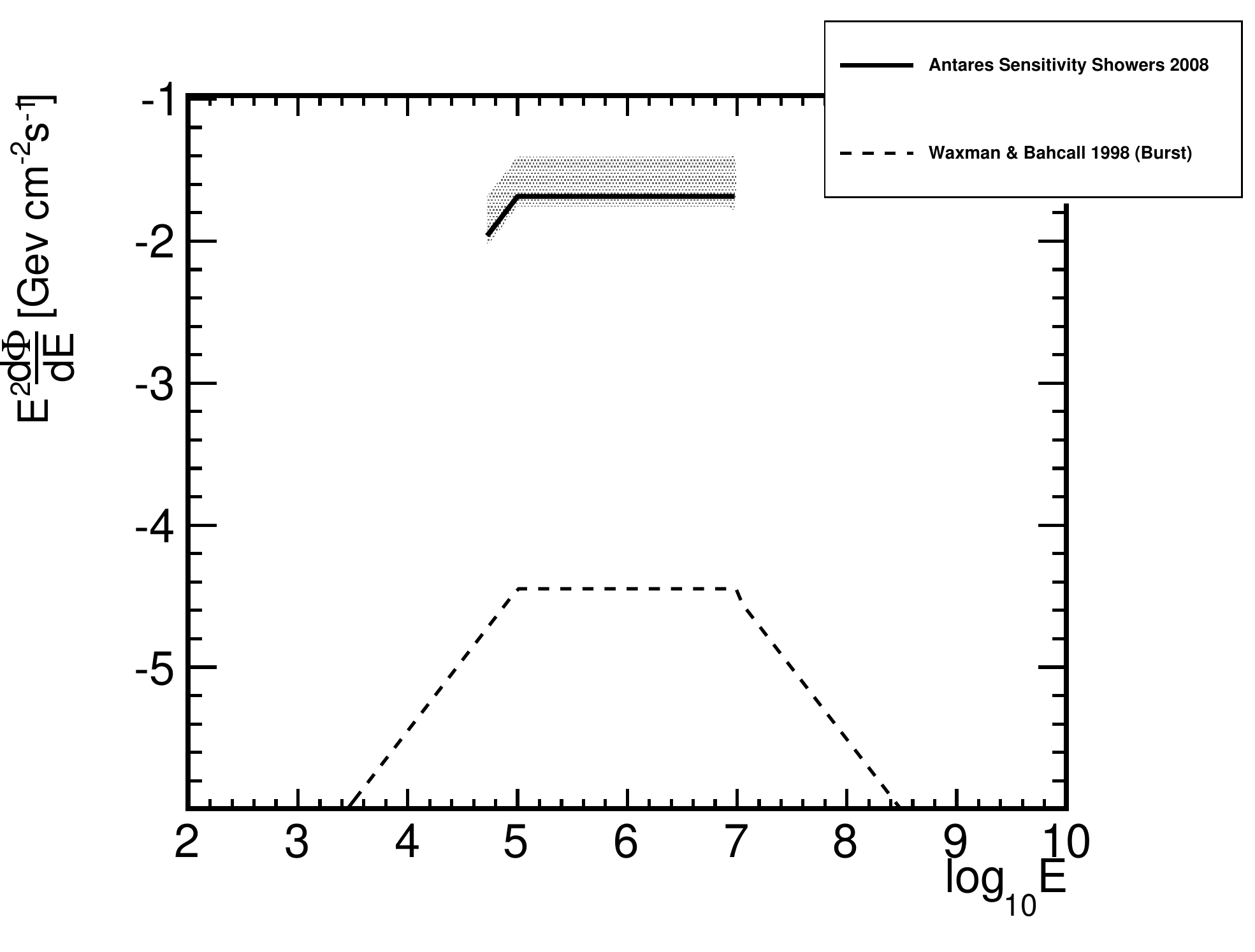}
\caption{Sensitivity of the Antares detector for neutrino induced showers from GRBs.}\label{fig:results_sensitivity} 
\end{figure}
The optimization of the quality cuts is made in order to maximize the discovery potential of the analysis. The optimized values lead to a $5\sigma$ discovery if 4 events are measured within the time window of the GRB ($100~\s$ in the optimization procedure) or a $3\sigma$ discovery when 3 events are measured. This method will then be applied for each observed GRB. The 90\% C.L. average upper limit of Antares (or sensitivity) for a neutrino induced shower in coincidence of  GRB trigger is shown in Figure \ref{fig:results_sensitivity}. The dashed line shows the model used for the estimate of the neutrino rate \cite{waxman1997}. The thick black line is the average upper limit on the neutrino flux for 2008 data, calculated using the Feldman-Cousins method \cite{FC1998}.
 The sensitivity of Antares for showers in coincidence with a GRB is $E_{\nu}^2\frac{d\Phi}{dE_{\nu}} \leq 1.67 \times 10^{-2} \textrm{GeV} \textrm{cm}^{-2} \textrm{s}^{-1} \; \textrm{for }10^5<E_{\nu}<10^7$. This is the sensitivity corresponding to the 2008 data set, during which the average background rate was $1.782 \x 10^{-4}$ Hz.




\bibliographystyle{aipproc}   

\bibliography{bibliography}

\IfFileExists{\jobname.bbl}{}
 {\typeout{}
  \typeout{******************************************}
  \typeout{** Please run "bibtex \jobname" to optain}
  \typeout{** the bibliography and then re-run LaTeX}
  \typeout{** twice to fix the references!}
  \typeout{******************************************}
  \typeout{}
 }

\end{document}